# Nonuniform Sparse Recovery with Subgaussian Matrices

Ulaş Ayaz and Holger Rauhut

Hausdorff Center for Mathematics and Institute for Numerical Simulation
University of Bonn
Endenicher Allee 60, 53115 Bonn, Germany
{ulas.ayaz,rauhut}@hcm.uni-bonn.de

October 30, 2018

**Abstract**

Compressive sensing predicts that sufficiently sparse vectors can be recovered from highly incomplete information using efficient recovery methods such as $\ell_1$-minimization. Random matrices have become a popular choice for the measurement matrix. Indeed, near-optimal uniform recovery results have been shown for such matrices. In this note we focus on nonuniform recovery using subgaussian random matrices and $\ell_1$-minimization. We provide conditions on the number of samples in terms of the sparsity and the signal length which guarantees that a fixed sparse signal can be recovered with a random draw of the matrix using $\ell_1$-minimization. Our proofs are short and provide explicit and good constants.

## 1 Introduction

Compressive sensing allows to reconstruct signals from far fewer measurements than what was considered necessary before. The seminal papers by E. Candes, J. Romberg, T. Tao [5,7] and by D. Donoho [11] have triggered a large research activity in mathematics, engineering and computer science with a lot of potential applications.

In mathematical terms we aim at solving the linear system of equations $y = Ax$ for $x \in \mathbb{C}^N$ when $y \in \mathbb{C}^m$ and $A \in \mathbb{C}^{m \times N}$ are given, and when $m \ll N$. Clearly, in general this task is impossible since even if $A$ has full rank then there are infinitely many solutions to this equation. The situation dramatically changes if $x$ is sparse, that is, $\|x\|_0 := \#\{\ell, x_\ell \neq 0\}$ is small. We note that $\|\cdot\|_0$ is called $\ell_0$-norm although it is not a norm.

As a first approach one is led to solve the optimization problem

$$\min_{z \in \mathbb{C}^N} \|z\|_0 \text{ subject to } Az = y, \tag{1.1}$$

where $Ax = y$. Unfortunately, this problem is NP-hard in general, so intractable in practice. It has become common to replace the $\ell_0$-minimization problem by the $\ell_1$-minimization problem

$$\min_{z \in \mathbb{C}^N} \|z\|_1 \text{ subject to } Az = y, \tag{1.2}$$

where $Ax = y$. This problem can be solved by efficient convex optimization techniques [3]. As a key result of compressive sensing, under appropriate conditions on $A$ and on the sparsity of $x$, $\ell_1$-minimization indeed reconstructs the original $x$. There are basically two types of recovery results:

- **Uniform recovery**: Such results state that with high probability on the draw of the random matrix, *every* sparse vector can be reconstructed under appropriate conditions.

- **Nonuniform recovery**: Such results state that a given sparse vector $x$ can be reconstructed with high probability on the the draw of the matrix under appropriate conditions. The difference to uniform recovery is that nonuniform recovery does not imply that there is a matrix that recovers all $x$ simultaneously. Or in other words, the small exceptional set of matrices for which recovery fails may depend on $x$.

Uniform recovery via $\ell_1$-minimization is for instance satisfied if the by-now classical restricted isometry property (RIP) holds for $A$ with high probability [4,6]. A common choice is to take $A \in \mathbb{R}^{m \times N}$ as a Gaussian random matrix, that is, the entries of $A$ are independent normal distributed mean-zero random variables of variance 1. If

$$m \geq Cs \ln(N/s), \tag{1.3}$$

then with probability at least $1 - e^{-cm}$ we have uniform recovery of all $s$-sparse vectors $x \in \mathbb{R}^N$ using $\ell_1$-minimization and $A$ as measurement matrix, see e.g. [7, 13, 19].

In this note we consider nonuniform sparse recovery using Gaussian and more general subgaussian random matrices in connection with $\ell_1$-minimization. Our main results below provide non-uniform recovery guarantees with an explicit and good constant. In contrast to other works such as [12, 13] we can treat also the recovery of complex vectors. Moreover, we get also good constants in the subgaussian case, and in particular, for Bernoulli matrices.

## 2 Main results

### 2.1 Gaussian case

We say that an $m \times N$ random matrix $A$ is Gaussian if its entries are independent and standard normal distributed random variables, that is, having mean zero and variance 1. Our nonuniform sparse recovery result for Gaussian matrices and $\ell_1$-minimization reads as follows.

**Theorem 2.1.** *Let $x \in \mathbb{C}^N$ with $\|x\|_0 = s$. Let $A \in \mathbb{R}^{m \times N}$ be a randomly drawn Gaussian matrix, and let $\varepsilon \in (0, 1)$. If*

$$m \geq s \left[\sqrt{2 \ln(2N/\varepsilon) + 1} + \sqrt{2 \ln(2/\varepsilon)/s}\right]^2 \tag{2.1}$$

*then with probability at least $1 - \varepsilon$ the vector $x$ is the unique solution to the $\ell_1$-minimization problem* (1.2).

**Remark:** In the asymptotic regime $N, s \to \infty$, (2.1) becomes simply

$$m \geq 2s \ln(3N/\varepsilon). \tag{2.2}$$

Comparing with (1.3) we realize that the log-term falls slightly short of the optimal one $\log(N/s)$. However, we emphasize that our proof is short, and the constant is explicit and good. Indeed, when in addition $s/N \to 0$ then we nevertheless reveal the conditions found by Donoho and Tanner [12, 13], and in particular, the optimal constant 2. Note that Donoho and Tanner used methods from random polytopes, which are quite different from our proof technique.

## 2.2 Subgaussian case

We generalize our recovery result for matrices with entries that are independent subgaussian random variables. A random variable $X$ is called *subgaussian* if there are constants $\beta, \theta > 0$ such that
$$\mathbb{P}(|X| \geq t) \leq \beta e^{-\theta t^2} \quad \text{for all } t > 0. \tag{2.3}$$
It can be shown [22] that $X$ is subgaussian with $\mathbb{E}X = 0$ if and only if there exists a constant $c$ (depending only on $\beta$ and $\theta$) such that
$$\mathbb{E}[\exp(\lambda X)] \leq e^{c\lambda^2} \quad \text{for all } \lambda \in \mathbb{R}. \tag{2.4}$$
Important special cases of subgaussian mean-zero random variables are standard Gaussians, and Rademacher (Bernoulli) variables, that is, random variables that take the values $\pm 1$ with equal probability. For both of these random variables the constant $c = 1/2$, see also Section 2.3.

A random matrix with entries that are independent mean-zero subgaussian random variables with the same constant $c$ in (2.4) is called a subgaussian random matrix. Note that the entries are not required to be identically distributed.

**Theorem 2.2.** *Let $x \in \mathbb{C}^N$ with $\|x\|_0 = s$. Let $A \in \mathbb{R}^{m \times N}$ be a random draw of a subgaussian matrix with constant $c$ in (2.4), and let $\varepsilon \in (0,1)$. If*
$$m \geq \frac{4c}{1 - \left(\frac{3C}{4c}\right)^{1/2} \ln(4N/\varepsilon)^{-1/2} \ln(4/\varepsilon)^{1/2}} s \ln(4N/\varepsilon), \tag{2.5}$$
*(where we assume additionally that $N, \varepsilon$ are such that the denominator above is positive) then with probability at least $1 - \varepsilon$ the vector $x$ is the unique solution to the $\ell_1$-minimization problem (1.2). The constant $C$ in (2.5) only depends on $c$.*

More precisely, the constant $C = 1.646 \tilde{c}^{-1}$, where $\tilde{c} = \tilde{c}(c)$ is the constant from Lemma (E.1) below, see also Lemma E.2.

**Remark:** If we consider the asymptotic regime $N \to \infty$, the number of measurements that guarantees recovery with high probability scales like $(4c) s \ln(N)$.

## 2.3 Bernoulli case

We specialize the previous result for subgaussian matrices to Bernoulli (Rademacher) matrices, that is, random matrices with independent entries taking the value $\pm 1$ with equal probability. We are then able to give explicit constants for the constants appearing in the result of Theorem 2.2. If $Y$ is a Bernoulli random variable, then
$$\mathbb{E}(\exp(\lambda Y)) = \frac{1}{2}\left(e^\lambda + e^{-\lambda}\right) \leq e^{\frac{1}{2}\lambda^2}$$
The last inequality can be derived by using Taylor series. This shows that the subgaussian constant $c = 1/2$ in the Bernoulli case. Further, we have the following concentration inequality for a matrix $B \in \mathbb{R}^{m \times N}$ with entries as independent realizations of $\pm 1/\sqrt{m}$,
$$\mathbb{P}\left(\left|\|\tilde{B}x\|_2^2 - \|x\|_2^2\right| > t\|x\|_2^2\right) \leq 2 e^{-\frac{m}{2}(t^2/2 - t^3/3)}, \tag{2.6}$$
for all $x \in \mathbb{R}^N$, $t \in (0,1)$, see e.g. [1, 2]. We can simply estimate $t^3 < t^2$ in (2.6) and get $\tilde{c} = 1/12$ in Lemma E.2 and consequently $C = 19.76$.

**Corollary 2.3.** *Let $x \in \mathbb{C}^N$ with $\|x\|_0 = s$. Let $A \in \mathbb{R}^{m \times N}$ be a matrix with entries that are independent Bernoulli random variables, and let $\varepsilon \in (0,1)$. If*

$$m \geq \frac{2}{1 - 5.45\ln(4N/\varepsilon)^{-1/2}\ln(4/\varepsilon)^{1/2}} s\ln(4N/\varepsilon), \qquad (2.7)$$

*then with probability at least $1-\varepsilon$ the vector $x$ is the unique solution to the $\ell_1$-minimization problem* (1.2).

### 2.4 Relation to previous work

Recently, there have been several papers dealing with nonuniform recovery. Most of these papers only consider the Gaussian case while our results extend to subgaussian and in particular to Bernoulli matrices.

As already mentioned, Donoho and Tanner [13] obtain nonuniform recovery results (terminology is "weak phase transitions") for Gaussian matrices via methods from random polytopes. They operate essentially in an asymptotic regime (although some of their results apply also for finite values of $N, m, s$). They consider the case that

$$m/N \to \delta,\ s/m \to \rho,\ \log(N)/m \to 0,\ N \to \infty,$$

where $\rho, \delta$ are some fixed values. Recovery conditions are then expressed in terms of $\rho$ and $\delta$ in this asymptotic regime. In particular, they get a (weak) transition curve $\rho_W(\delta)$ such that $\rho < \rho_W(\delta)$ implies recovery with high probability and $\rho > \rho_W(\delta)$ mean failure with high probability (as $N \to \infty$). Moreover, they show that $\rho_W(\delta) \sim 2\log(\delta^{-1})$ as $\delta \to 0$. Translated back into the quantities $N, m, s$ this gives $m \geq 2s\log(N)$ in an asymptotic regime, which is essentially (2.2).

Candès and Plan give a rather general framework for nonuniform recovery in [8], which applies to measurement matrices with independent rows having bounded entries. In fact, they prove a recovery condition for such random matrices of the form $m \geq Cs\ln(N)$ for some constant $C$. However, they do not get explicit and good constants. Dossal et al. [14], derive a recovery condition for Gaussian matrices of the form $m \geq cs\ln(N)$, where $c$ approaches 2 in an asymptotic regime. These both papers also obtain stability results for noisy measurements.

Finally, Chandrasekaran et al. [9] use convex geometry in order to obtain nonuniform recovery results. They develop a rather general framework that applies also to low rank recovery and further setups. However, they can only treat Gaussian measurements. They approach the recovery problem via Gaussian widths of certain convex sets. In particular, they estimate the number of Gaussian measurements needed in order to recover an $s$ sparse vector by $m \geq 2s(\ln(p/s - 1) + 1)$ which is essentially the optimal result. It is not straightforward to extend their method to subgaussian measurements as they heavily use the rotation invariance of Gaussian random vectors.

## 3 Proofs

### 3.1 Notation

We start with setting up some notation needed in the proofs. Let $[N]$ denote the set $\{1, 2, \ldots, N\}$. The column submatrix of a matrix $A$ consisting of the columns indexed by $S$ is written $A_S = (a_j)_{j \in S}$ where $S \subset [N]$ and $a_j \in \mathbb{R}^m$, $j = 1, \ldots, m$ denote the columns of $A$. Similarly $x^S \in \mathbb{C}^S$ denotes the vector $x \in \mathbb{C}^N$ restricted to the entries in $S$, and $x \in \mathbb{C}^N$ is called $s$-sparse if $\text{supp}(x) = \{\ell : x_\ell \neq 0\} = S$ with $S \subset [N]$ and $|S| = s$, i.e., $\|x\|_0 = s$. We further need to introduce the sign vector $\text{sgn}(x) \in \mathbb{C}^N$ having entries

$$\text{sgn}(x)_j := \begin{cases} \frac{x}{|x_j|} & \text{if } x_j \neq 0, \\ 0 & \text{if } x_j = 0, \end{cases} \quad j \in [N].$$

The Moore-Penrose pseudo-inverse of a matrix $B$ such that $(B^*B)$ is invertible is given by $B^\dagger = (B^*B)^{-1}B^*$, so that $B^\dagger B = \text{Id}$, where Id is the identity matrix.

## 3.2 The Gaussian case

We set $S := \text{supp}(x)$, which has a cardinality $s$. By Corollary A.2, for recovery via $\ell_1$-minimization, it is sufficient to show that

$$|\langle (A_S)^\dagger a_\ell, \text{sgn}(x^S) \rangle| = |\langle a_\ell, (A_S^\dagger)^* \text{sgn}(x^S) \rangle| < 1 \text{ for all } \ell \in [N] \setminus S.$$

Therefore, the failure probability for recovery is bounded by

$$\mathcal{P} := \mathbb{P}(\exists \ell \notin S \ |\langle (A_S)^\dagger a_\ell, \text{sgn}(x^S) \rangle| \geq 1).$$

If we condition $X := \langle a_\ell, (A_S^\dagger)^* \text{sgn}(x^S) \rangle$ on $A_S$, it is a Gaussian random variable. Further, $X = \sum_{j=1}^m (a_\ell)_j [(A_S^\dagger)^* \text{sgn}(x^S)]_j$ is centered so its variance $\nu^2$ can be estimate by

$$\nu^2 = \mathbb{E}(X^2) = \sum_{j=1}^m \mathbb{E}[(a_\ell)_j^2][(A_S^\dagger)^* \text{sgn}(x^S)]_j^2$$

$$= \|(A_S^\dagger)^* \text{sgn}(x^S)\|_2^2 \leq \sigma_{\min}^{-2}(A_S) \|\text{sgn}(x^S)\|_2^2 = \sigma_{\min}^{-2}(A_S) s,$$

where $\sigma_{\min}$ denotes the smallest singular value. The last inequality uses the fact that $\|(A_S^\dagger)^*\|_{2 \to 2} = \|A_S^\dagger\|_{2 \to 2} = \sigma_{\min}^{-1}(A_S)$. Then it follows that

$$\mathcal{P} \leq \mathbb{P}\left(\exists \ell \notin S \ |\langle (A_S)^\dagger a_\ell, \text{sgn}(x^S) \rangle| \geq 1 \ \Big| \ \|(A_S^\dagger)^* \text{sgn}(x^S)\|_2 < \alpha\right)$$
$$+ \mathbb{P}(\|(A_S^\dagger)^* \text{sgn}(x^S)\|_2 \geq \alpha)$$
$$\leq 2N \exp(-1/2\alpha^2) + \mathbb{P}(\sigma_{\min}^{-1}(A_S) \sqrt{s} \geq \alpha). \tag{3.1}$$

The inequality in (3.1) uses the tail estimate (C.1) for a gaussian random variable, the union bound, and the independence of $a_\ell$ and $A_S$. The first term in (3.1) is bounded by $\varepsilon/2$ if

$$\alpha \leq \frac{1}{\sqrt{2 \ln(2N/\varepsilon)}}. \tag{3.2}$$

The second term in (3.1) can be estimated using (B.1) below,

$$\mathbb{P}(\sigma_{\min}^{-1}(A_S) \sqrt{s} \geq \alpha)$$
$$= \mathbb{P}(\sigma_{\min}(A_S) \leq \sqrt{s}/\alpha) = \mathbb{P}\left(\sigma_{\min}(A_S/\sqrt{m}) \leq \sqrt{s}/(\sqrt{m}\alpha)\right)$$
$$\leq \exp\left(\frac{-m(1 - (\alpha^{-1} + 1)\sqrt{s/m})^2}{2}\right). \tag{3.3}$$

If we choose $\alpha$ that makes (3.2) an equality, plug it into condition (3.3), and require that (3.3) is bounded by $\varepsilon/2$ we arrive at the condition

$$m \geq s \left[\sqrt{2\ln(2N/\varepsilon)+1} + \sqrt{2\ln(2/\varepsilon)/s}\right]^2,$$

which ensures recovery with probability at least $1-\varepsilon$. This concludes the proof of Theorem 2.1. □

### 3.3 Subgaussian case

We follow a similar path as in the proof of Gaussian case. We denote $S := \mathrm{supp}(x)$. We can bound the failure probability $\mathcal{P}$ by

$$\mathcal{P} \leq \mathbb{P}\left(\exists \ell \notin S \ |\langle (A_S)^\dagger a_\ell, \mathrm{sgn}(x^S)\rangle| \geq 1 \,\Big|\, \|(A_S^\dagger)^* \mathrm{sgn}(x^S)\|_2 < \alpha\right)$$
$$+ \mathbb{P}(\|(A_S^\dagger)^* \mathrm{sgn}(x^S)\|_2 \geq \alpha). \tag{3.4}$$

The first term in (3.4) can be bounded by using Lemma D.1. Conditioning on $A_S$ and $\|(A_S^\dagger)^* \mathrm{sgn}(x^S)\|_2 < \alpha$ we get

$$\mathbb{P}(|\langle (A_S)^\dagger a_\ell, \mathrm{sgn}(x^S)\rangle| \geq 1) = \mathbb{P}(|\sum_{j=1}^m (a_\ell)_j [(A_S^\dagger)^* \mathrm{sgn}(x^S)]_j| \geq 1) \leq 2\exp(-1/(4c\alpha^2)).$$

So by the union bound the first term in (3.4) can be estimated by $2N\exp(-1/(4c\alpha^2))$, which in turn is no larger than $\varepsilon/2$ provided

$$\alpha \leq \sqrt{1/(4c\ln(4N/\varepsilon))}. \tag{3.5}$$

For the second term in (3.4), we have

$$\mathbb{P}(\|(A_S^\dagger)^* \mathrm{sgn}(x^S)\|_2 \geq \alpha) \leq \mathbb{P}(\sigma_{\min}^{-1}(A_S)\sqrt{s} \geq \alpha)$$
$$= \mathbb{P}(\sigma_{\min}(A_S) \leq \sqrt{s}/\alpha) = \mathbb{P}\left(\sigma_{\min}(A_S/\sqrt{m}) \leq \frac{1}{\sqrt{m}}\frac{\sqrt{s}}{\alpha}\right).$$

Lemma E.1 and Lemma E.2 imply that a matrix $B := A_S/\sqrt{m}$ with normalized subgaussian rows satisfy

$$\mathbb{P}(\sigma_{\min}(B) < \sqrt{1-\delta}) < \mathbb{P}(\|B^*B - \mathrm{Id}\|_{2\to 2} \geq \delta) < \varepsilon/2$$

provided $m \geq C\delta^{-2}(3s + \ln(4\varepsilon^{-1}))$, where $C$ depends on subgaussian constant $c$. The choice $\frac{1}{\sqrt{m}}\frac{\sqrt{s}}{\alpha} = \sqrt{1-\delta}$ yields $\delta = 1 - \frac{s}{m\alpha^2}$. Combining these arguments and choosing $\alpha$ that makes (3.5) an equality, we can bound the failure probability by $\varepsilon$ provided

$$m \geq C\left(1 - \frac{4cs\ln(4N/\varepsilon)}{m}\right)^{-2}(3s + \ln(4/\varepsilon)). \tag{3.6}$$

We define the variable $\gamma := 1 - \left(\frac{3C}{4c}\right)^{1/2}\ln(4N/\varepsilon)^{-1/2}\ln(4/\varepsilon)^{1/2}$. Observe that $\gamma \in (0,1)$ for $N$ large enough. If $\gamma \geq 4cs\ln(4N/\varepsilon)/m$, that is, if

$$m \geq \frac{4c}{\gamma}s\ln(4N/\varepsilon), \tag{3.7}$$

then condition (3.6) is implied by

$$m \geq 3C(1-\gamma)^{-2}(s + \ln(4/\varepsilon)/3). \tag{3.8}$$

If we plug $\gamma$ into (3.7) and (3.8), it can be seen that (3.7) implies (3.8). This completes the proof of Theorem 2.2. □

## A  Recovery conditions

In this section we state some theorems that were used in the proof of main theorem, directly or indirectly. The proofs of Theorems 2.1 and 2.2 require a condition for sparse recovery, which not only depends on the matrix $A$ but also on the sparse vector $x \in \mathbb{C}^N$ to be recovered. The following theorem is due to J.J. Fuchs [15] in the real-valued case and was extended to the complex case by J. Tropp [21], see also [20, Theorem 2.8] for a slightly simplified proof.

**Theorem A.1.** *Let $A \in \mathbb{C}^{m \times N}$ and $x \in \mathbb{C}^N$ with $S := \mathrm{supp}(x)$. Assume that $A_S$ is injective and that there exists a vector $h \in \mathbb{C}^m$ such that*

$$A_S^* h = \mathrm{sgn}(x^S),$$
$$|(A^*h)_\ell| < 1, \ \ell \in [N] \setminus S.$$

*Then $x$ is the unique solution to the $\ell_1$-minimization problem (1.2) with $Ax = y$.*

Choosing the vector $h = (A^\dagger)^* \mathrm{sgn}(x^S)$ leads to the following corollary.

**Corollary A.2.** *Let $A \in \mathbb{C}^{m \times N}$ and $x \in \mathbb{C}^N$ with $S := \mathrm{supp}(x)$. If the matrix $A_S$ is injective and if*

$$|\langle (A_S)^\dagger a_\ell, \mathrm{sgn}(x^S) \rangle| < 1 \ \text{ for all } \ \ell \in [N] \setminus S$$

*then the vector $x$ is the unique solution to the $\ell_1$-minimization problem (1.1) with $y = Ax$.*

## B  Singular values of Gaussian matrix

An elegant estimation for the smallest singular value of a normalized Gaussian matrix $B \in \mathbb{R}^{m \times s}$, where the entries of $B$ are independent and follow the normal distribution $\mathcal{N}(0, 1/m)$, was provided in [10],

$$\mathbb{P}(\sigma_{\min}(B) < 1 - \sqrt{s/m} - r) \leq e^{-mr^2/2}. \tag{B.1}$$

Its proof relies on the Slepian-Gordon Lemma [16, 17] and concentration of measure for Lipschitz functions [18].

## C  Tail estimate for a gaussian random variable

For a mean-zero Gaussian random variable $X$ with variance $\sigma^2$ we have the tail estimate

$$\mathbb{P}(|X| > t) \leq e^{-t^2/2\sigma^2}. \tag{C.1}$$

Indeed, a mean-zero Gaussian variable $g$ with variance satisfies by [20, Lemma 10.2]

$$\mathbb{P}(|g| > t) = \frac{2}{\sqrt{2\pi}} \int_t^\infty e^{-t^2/2} dt \leq e^{-t^2/2}.$$

Rescaling gives the tail estimate (C.1).

# D  Tail estimate for sums of subgaussian variables

The following estimate for sums of subgaussian random variables appears for instance in [22].

**Lemma D.1.** *Let $X_1, \ldots, X_M$ be a sequence of independent mean-zero subgaussian random variables with the same parameter $c$ as in (2.4). Let $a \in \mathbb{R}^M$ be some vector. Then $Z := \sum_{j=1}^M a_j X_j$ is subgaussian, that is, for $t > 0$,*

$$\mathbb{P}(|\sum_{j=1}^M a_j X_j| \geq t) \leq 2\exp(-t^2/(4c\|a\|_2^2)).$$

*Proof.* For convenience we provide a proof. By independence we have

$$\mathbb{E}\exp(\theta \sum_{j=1}^M a_j X_j) = \mathbb{E}\prod_{i=1}^M \exp(\theta a_j X_j) = \prod_{i=1}^M \mathbb{E}\exp(\theta a_j X_j) \leq \prod_{i=1}^M \exp(\theta a_j X_j)$$
$$= \exp(c\|a\|_2^2 \theta^2).$$

This shows that $Z$ subgaussian with parameter $c\|a\|_2^2$ in (2.4). We apply Markov's inequality to get

$$\mathbb{P}(Z \geq t) = \mathbb{P}(\exp(\theta Z) \geq \exp(\theta t)) \leq \mathbb{E}[\exp(\theta Z)]e^{-\theta t} \leq e^{c\|a\|_2^2 \theta^2 - \theta t}.$$

The optimal choice $\theta = t/(2c\|a\|_2^2)$ yields

$$\mathbb{P}(Z \geq t) \leq e^{-t^2/(4c\|a\|_2^2)}.$$

Repeating the above computation with $-Z$ instead of $Z$ shows that

$$\mathbb{P}(-Z \geq t) \leq e^{-t^2/(4c\|a\|_2^2)},$$

and the union bound yields the desired estimate $\mathbb{P}(|Z| \geq t) \leq 2e^{-t^2/(4c\|a\|_2^2)}$. □

# E  Concentration Inequalities

The following concentration inequality for subgaussian random variables appears, for instance, in [1, 19].

**Lemma E.1.** *Let $A$ be an $m \times N$ random matrix with independent, isotropic, and subgaussian rows with the same parameter $c$ as in (2.4). Then, for all $x \in \mathbb{R}^N$ and every $t \in (0,1)$, normalized matrix $\tilde{A} = \frac{1}{\sqrt{m}} A$ satisfies*

$$\mathbb{P}(|\|\tilde{A}x\|_2^2 - \|x\|_2^2| > t\|x\|_2^2) \leq 2\exp(-\tilde{c}mt^2), \tag{E.1}$$

*where $\tilde{c}$ depends only on $c$.*

Combing the above concentration inequality with the net technique we can derive the following estimate on the condition of (submatrices of) subgaussian random matrices. While this is well-known in principle the right scaling in $\delta$ seemingly has not appeared elsewhere in the literature, compare with [2, 19].

**Lemma E.2.** *Let $S \subset [N]$ with $\operatorname{card}(S) = s$. Suppose that $m \times N$ random matrix $A$ is drawn according to a probability distribution for which the concentration inequality (E.1) holds, that is, for $t > 0$,*

$$\mathbb{P}(|\|Ax\|_2^2 - \|x\|_2^2| > t\|x\|_2^2) \leq 2\exp(-\tilde{c}mt^2) \text{ for all } x \in \mathbb{R}^N,$$

*for some $\tilde{c} \in \mathbb{R}$. Then, for $\delta \in (0,1)$,*

$$\|A_S^* A_S - \operatorname{Id}\|_{2 \to 2} \leq \delta$$

*with probability at least $1 - \varepsilon$ provided*

$$m \geq C\delta^{-2}(3s + \ln(2\varepsilon^{-1})), \tag{E.2}$$

*with $C = 1.646\tilde{c}^{-1}$.*

*Proof.* Since most available statements have an additional $\log(\delta^{-1})$-term in (E.2), we include the proof of this lemma for the sake of completeness.

Let $\rho \in (0, \sqrt{2}-1)$ be a number to be determined later. According to a classical covering number argument, see e.g. [20, Proposition 10.1], there exists a finite subset $U$ of the unit sphere $\mathcal{S} = \{x \in \mathbb{R}^N, \operatorname{supp}(x) \subset S, \|x\|_2 = 1\}$, which satisfies

$$|U| \leq \left(1 + \frac{2}{\rho}\right)^s \quad \text{and} \quad \min_{u \in U} \|z - u\|_2 \leq \rho \quad \text{for all } z \in \mathcal{S}.$$

The concentration inequality (E.1) yields

$$\mathbb{P}\left(|\|Au\|_2^2 - \|u\|_2^2| > t\|u\|_2^2 \text{ for some } u \in \mathcal{U}\right)$$
$$\leq \sum_{u \in U} \mathbb{P}\left(|\|Au\|_2^2 - \|u\|_2^2| > t\|u\|_2^2\right) \leq 2|U|\exp\left(-\tilde{c}t^2 m\right)$$
$$\leq 2\left(1 + \frac{2}{\rho}\right)^s \exp\left(-\tilde{c}t^2 m\right).$$

The positive number $t$ will be set later depending on $\delta$ and on $\rho$. Let us assume for now that the realization of the random matrix $A$ yields

$$\left|\|Au\|_2^2 - \|u\|_2^2\right| \leq t \qquad \text{for all } u \in U. \tag{E.3}$$

By the above, this occurs with probability exceeding

$$1 - 2\left(1 + \frac{2}{\rho}\right)^s \exp\left(-\tilde{c}t^2 m\right). \tag{E.4}$$

Next we show that (E.3) implies $|\|Ax\|_2^2 - \|x\|_2^2| \leq \delta$ for all $x \in \mathcal{S}$, that is $\|A_S^* A_S - \operatorname{Id}\|_{2 \to 2} \leq \delta$ (when $t$ is determined appropriately). Let $B = A_S^* A_S - \operatorname{Id}$, so that we have to show $\|B\|_{2 \to 2} \leq \delta$. Note that (E.3) means that $|\langle Bu, u \rangle| \leq t$ for all $u \in U$. Now consider a vector $x \in \mathcal{S}$, for which we choose a vector $u \in U$ satisfying $\|x - u\|_2 \leq \rho < \sqrt{2} - 1$. We obtain

$$|\langle Bx, x \rangle| = |\langle B(u + x - u), u + x - u \rangle|$$
$$= |\langle Bu, u \rangle + \langle B(x-u), x-u \rangle + 2\langle Bu, x-u \rangle|$$
$$\leq |\langle Bu, u \rangle| + |\langle B(x-u), x-u \rangle| + 2\|Bu\|_2 \|x-u\|_2$$
$$\leq t + \|B\|_{2 \to 2} \rho^2 + 2\|B\|_{2 \to 2} \rho.$$

Taking the supremum over all $x \in \mathcal{S}$, we deduce that

$$\|B\|_{2\to 2} \le t + \|B\|_{2\to 2}\left(\rho^2 + 2\rho\right), \qquad \text{i.e.,} \qquad \|B\|_{2\to 2} \le \frac{t}{2-(\rho+1)^2}.$$

Note that the division by $2-(\rho+1)^2$ is justified by the assumption that $\rho < \sqrt{2}-1$. Then we choose

$$t = t_{\delta,\rho} := \left(2-(\rho+1)^2\right)\delta,$$

so that $\|B\|_{2\to 2} \le \delta$, and with our definition of $t$,

$$\mathbb{P}\left(\|A_S^* A_S - \mathrm{Id}\|_{2\to 2} > \delta\right) \le 2\left(1+\frac{2}{\rho}\right)^s \exp\left(-\tilde{c}\delta^2(2-(\rho+1)^2)^2 m\right). \tag{E.5}$$

Hence, $\|A_S^* A_S - \mathrm{Id}\|_{2\to 2} \le \delta$ with probability at least $1-\varepsilon$ provided

$$m \ge \frac{1}{\tilde{c}(2-(\rho+1)^2)^2}\delta^{-2}\left(\ln(1+2/\rho)s + \ln(2\varepsilon^{-1})\right). \tag{E.6}$$

Now we choose $\rho$ such that $\ln(1+2/\rho) = 3$, that is, $\rho = 2/(e^3-1)$. Then (E.6) gives the condition

$$m \ge C\delta^{-2}\left(3s + \ln(2\varepsilon^{-1})\right) \tag{E.7}$$

with $C = 1.646\,\tilde{c}^{-1}$. This concludes the proof. $\square$

## Acknowledgement

The authors would like to thank the Hausdorff Center for Mathematics for support, and acknowledge funding through the WWTF project SPORTS (MA07-004). We would also like to thank Simon Foucart. Indeed, the proof of Lemma E.2 is taken from a book draft that the second author is currently preparing with him.